\documentclass[aps,prd,11pt,twoside,tightenlines,nofootinbib,showpacs,preprint,superscriptaddress]{revtex4-1}
\usepackage[colorlinks=true]{hyperref}
\usepackage{xcolor}
\hypersetup{colorlinks=true, citecolor=blue, urlcolor=blue, linkcolor=blue}
\usepackage{amsmath,amssymb}
\usepackage[final]{graphicx}
\usepackage{subcaption}
\usepackage{float}
\usepackage[toc,page]{appendix}

\begin{document}

\title{Study of $D_{s}^{+} \rightarrow K^{+} K^{-} \pi^{+}$ decay}

\author{Zhong-Yu Wang}
\affiliation{School of Physics and Electronics, Hunan Key Laboratory of Nanophotonics and Devices, Central South University, Changsha 410083, China}

\author{Jing-Yu Yi}
\affiliation{School of Physics and Electronics, Hunan Key Laboratory of Nanophotonics and Devices, Central South University, Changsha 410083, China}

\author{Zhi-Feng Sun}
\affiliation{School of Physical Science and Technology, Lanzhou University, Lanzhou 730000, China}
\affiliation{Research Center for Hadron and CSR Physics, Lanzhou University 
and Institute of Modern Physics of CAS, Lanzhou 730000, China}
\affiliation{Lanzhou Center for Theoretical Physics, Key Laboratory of Theoretical Physics of Gansu Province,
and Frontiers Science Center for Rare Isotopes, Lanzhou University, Lanzhou 730000, China}

\author{C. W. Xiao}
\email{xiaochw@csu.edu.cn}
\affiliation{School of Physics and Electronics, Hunan Key Laboratory of Nanophotonics and Devices, Central South University, Changsha 410083, China}

\date{\today}

\begin{abstract}
	
We investigate the $D_{s}^{+} \rightarrow K^{+} K^{-} \pi^{+}$ decay theoretically with the final state interactions, which is based on the chiral unitary approach and takes into account the external and internal $W$-emission mechanisms at the quark level. Only considering three resonances contributions, the $f_0(980)$ in $S$-wave, the $\bar {K}^{*}(892)^{0}$ and $\phi(1020)$ in $P$-wave, one can make a good description of the recent experimental data from BESIII Collaboration, where the contribution from $S$-wave is found to be small. Besides, we also make a calculation of the corresponding branching fractions, which are consistent with the results of BESIII Collaboration and Particle Data Group.

\end{abstract}

\maketitle

\section{Introduction}

The weak decays of heavy mesons have drawn much attention in experiments for understanding the hadronic decay properties and the resonance productions of the decay processes. In particular, some non-leptonic three-body weak decays of $D_{(s)}$ mesons  were measured, where many resonances were produced in the final state interactions. The Dalitz plot analysis for the decay of $D_s^+ \to \pi^+ \pi^- \pi^+$ was performed in Ref.~\cite{Aubert:2008ao}, where the resonance contributions from $f_0(980)$, $f_0(1370)$, $f_0(1500)$, etc, were considered, and it was found that the contributions from $S$-wave resonances in $\pi^+ \pi^-$ channel were dominant
\footnote{These findings was confirmed by the recent measurements of BESIII Collaboration~\cite{BESIII:2021jnf}.}. 
The doubly Cabibbo-suppressed decay $D_s^+ \to K^+ K^+ \pi^-$ was firstly reported in Ref.~\cite{Ko:2009tc}, where the branching ratios for corresponding channels were measured. The resonance contributions from $S$- and $P$-waves, such as $f_0(980)$, $\bar{K}^*(892)^{0}$, $\phi(1020)$, etc, were analyzed in Ref.~\cite{delAmoSanchez:2010yp} for the decay $D_s^+ \to K^+ K^- \pi^+$. Note that its decay fractions for different decay modes were consistent with the former measurements of the E687~\cite{Frabetti:1995sg} and CLEO~\cite{Mitchell:2009aa} Collaborations. In Ref.~\cite{Aaij:2018hik}, the branching fractions were measured precisely for the doubly Cabibbo-suppressed decays $D^+ \to K^-  K^+ K^+$, $D^+ \to  \pi^-  \pi^+ K^+$ and $D_s^+ \to K^+ K^+ \pi^-$ compared to the normal decays $D^+ \to K^- \pi^+ \pi^+$ and $D_s^+ \to K^- K^+ \pi^+$. Ref.~\cite{Ablikim:2019pit} reported the first observation of the $W$-annihilation dominant decays $D_s^+ \to a_0(980)^+ \pi^0$ and $D_s^+ \to a_0(980)^0 \pi^+$ in the analysis of the reaction $D_s^+ \to \pi^+ \pi^0 \eta$, where the effect of $a_0(980)^0-f_0(980)$ mixing was found to be negligible. For the doubly Cabibbo-suppressed decay $D^+ \to K^- K^+ K^+$, Ref.~\cite{Aaij:2019lwx} investigated the resonant contributions in detail and found that the dominant contributions came from $S$-wave component of the $K^-K^+$ system, which observed a large contribution of the $a_0(980)$ with the destructive interference from the $f_0(980)$. In Ref.~\cite{Ablikim:2019ibo}, the singly Cabibbo-suppressed decays $D^+ \to \eta \eta \pi^+$ and $D^{+(0)} \to \eta \pi^+ \pi^{0(-)}$ were observed and the corresponding absolute branching fractions were measured. 

On the other hand, the non-leptonic three-body weak decays of $D_{(s)}$ catch much theoretical attention~\cite{Chau:1982da,Chau:1987tk}. The Cabibbo-favored charmed-meson decay $D^+ \to K^- \pi^+\pi^+$ was investigated in Ref.~\cite{Niecknig:2015ija} using dispersion theory based on  the Khuri-Treiman formalism, where the final state interactions between all three decay products were explained well by the constarints from analyticity and unitarity. In this framework, more higher partial waves were taken into account, and it was found that the contribution from $K\pi$ components in $D$-wave  was small. With the same framework, a further study on the decays $D^+ \to K^- \pi^+\pi^+$ and $D^+ \to \bar{K}^0 \pi^0\pi^+$ was done in Ref.~\cite{Niecknig:2017ylb}, which made a consistent description on the Daltz plot data from the CLEO, FOCUS and BESIII collaborations. Applying chiral effective Lagrangians to extract the interaction information of $K\bar{K}$, Ref.~\cite{Aoude:2018zty} studied the decay $D^+ \to K^- K^+ K^+$ with the isobar model and the coupled channel $K$-matrix approach, where the resonance contributions from $f_0(980)$ and $a_0(980)$ could be distinguished with different isospins of the two-body rescattering amplitudes. With the short-distance $W$-boson annihilation mechanism and the final state interaction of triangle rescattering process, Ref.~\cite{Yu:2021euw} discussed the $a_0(980)$ resonance contribution in the $D_s^+ \to a_0(980) \rho (\omega)$ decays and made predictions for their branching fractions. Using the chiral unitary approach (ChUA)~\cite{Oller:1997ti, Oset:1997it, Oller:1997ng, Kaiser:1998fi, Oller:2000fj, Oset:2008qh}, which characterizes by the Bethe-Salpeter (BS) equation in coupled channels, Ref.~\cite{Xie:2014tma} studied the resonance productions of $f_{0}(500)$, $f_{0}(980)$ and $a_{0}(980)$ in the final state interaction of weak decay $D^0 \to \bar{K}^0 \pi^+ \pi^-$ and $D^0 \to \bar{K}^0 \pi^0 \eta$, respectively. For the first observed $W$-annihilation decays $D_s^+ \to a_0(980)^+ \pi^0$ and $D_s^+ \to a_0(980)^0 \pi^+$~\cite{Ablikim:2019pit}, Ref. \cite{Molina:2019udw} proposed an internal $W$-emission mechanism to explain the process $D_{s}^{+} \to \pi^{+} \pi^{0} \eta$, where the resonances $a_{0}(980)^{+}$ and $a_{0}(980)^{0}$ were dynamically generated utilizing the coupled channel interaction of $K \bar K$ and $\pi \eta$ in the final state interactions. These $W$-annihilation decays~\cite{Ablikim:2019pit} were also explained in Ref.~\cite{Hsiao:2019ait} by the triangle rescattering mechanism of $\rho \pi \eta^{(\prime)}$, where the $\pi\eta$ invariant mass spectra were well described and the corresponding branching ratios were in good agreement with the experimental measurement. Furthermore, with the internal and external $W$-emission mechanisms instead of the $W$-annihilation, and considering the contributions from the $\rho^+$ and the intermediate $\rho^+ \eta$ and $K^*\bar{K}/K\bar{K}^*$ triangle diagrams, Ref.~\cite{Ling:2021qzl} obtained a good description of the experimental data for the decay $D_s^+ \to \pi^+ \pi^0 \eta$~\cite{Ablikim:2019pit}, where the $a_{0}(980)$ was produced in the final state interaction of $\pi \eta$. Analogously, applying the internal and external $W$-emission mechanisms, the single Cabibbo-suppressed $D^{+} \rightarrow \pi^{+} \pi^{0} \eta$ decay~\cite{Ablikim:2019ibo} was investigated in Ref.~\cite{Duan:2020vye} with the contribution of $a_0(980)$ in the two-body $\pi \eta$ final state interaction. The double Cabibbo-suppressed $D^{+} \rightarrow K^{-}K^{+}K^{+}$ decay~\cite{Aaij:2019lwx} was studied in Ref.~\cite{Roca:2020lyi}, which used N/D method to extend the applicability range up to 1.4 GeV and got the invariant mass distributions in good agreement with experimental data. Exploiting the external $W$-emission mechanism, Ref.~\cite{Toledo:2020zxj} studied the Cabibbo favoured decay $D^{0} \rightarrow K^{-}\pi^{+}\eta$ reported in Ref.~\cite{Belle:2020fbd} and described the experimental data well only with the contributions from $a_0(980)$ and $\kappa$ (or called $K^*_0(700)$ state). Moreover, how the two-body invariant mass distributions could be affected by the different weights of the internal and external $W$-emission mechanisms in the study of the reactions $D^+ \to \pi^+ \eta \eta$ and $D^+ \to \pi^+ \pi^0 \eta$~\cite{Ikeno:2021kzf}, where the future experimental data were expected to be helpful to pin down the reaction mechanisms and the contribution of the $a_0(980)$ resonance. And more discussions about non-leptonic three body decays of $D^{0}$ or $D_{(s)}^{+}$ based on the ChUA can be found in Ref.~\cite{Oset:2016lyh}.

Recently, the BESIII Collaboration reported the precise measurement of the branching fraction for the $D_{s}^{+} \to K^+ K^- \pi^+$ decay and obtained $\mathcal{B} (D_{s}^{+} \to K^{+} K^{-} \pi^{+})=(5.47\pm0.08\pm0.13)\%$~\cite{Ablikim:2020xlq}, which was consistent with the former measurements~\cite{delAmoSanchez:2010yp, Frabetti:1995sg, Mitchell:2009aa}. In Ref.~\cite{Ablikim:2020xlq}, performing amplitude analysis with high-statistics sample, the $K^{+} K^{-}$ invariant mass distribution was studied with a model-independent method. However, from the extracted $S$-wave lineshape of the $K^{+} K^{-}$ mass spectrum, the contributions from $a_0(980)$ and/or $f_{0}(980)$ were not identified in the low $K^{+} K^{-}$ mass region. This is the motivation of the present work, where we try to investigate that which contribution of the $f_{0}(980)$ and $a_{0}(980)$ resonances is dominant or unique for the decay $D_{s}^{+} \to K^+ K^- \pi^+$. In our former work~\cite{Ahmed:2020qkv}, the resonance contributions from the $f_0(980)$ and $f_0(500)$ were investigated in final state interactions of the decays $B^0_{(s)} \to \phi \pi^+ \pi^-$. Thus, in the present work, we also focus on the resonance contributions in the final state interactions. Based on the ChUA and using the external $W$-emission mechanism, the Cabibbo favored weak decays $D_{s}^{+} \rightarrow \pi^{+} \pi^{+} \pi^{-}$ and $D_{s}^{+} \rightarrow K^{+} K^{-} \pi^{+}$ were analyzed in Ref.~\cite{Dias:2016gou}, where the experimental data for the invariant mass distributions of $K^+ K^-$~\cite{delAmoSanchez:2010yp} and $\pi^+ \pi^-$~\cite{BaBar:2008nlp} was well described and the $f_0(980)$ resonance contribution was found in both the $K^+ K^-$ and $\pi^+ \pi^-$ invariant mass distributions. In a recent work of~\cite{Wang:2021naf}, the invariant mass distributions of $K^+ K^-$ from both BABAR~\cite{delAmoSanchez:2010yp} and BESIII~\cite{Ablikim:2020xlq} Collaborations' measurements were studied by both the external and internal $W$-emission mechanisms, and the dominant contribution from the $f_0(980)$ was found in the $K^+ K^-$ spectrum. Different from their works~\cite{Dias:2016gou,Wang:2021naf}, which only focused on the invariant mass distributions of $K^+ K^-$,  a full analysis of both $K \bar{K}$ and $K \pi$ mass spectra is done in our work including the $S$- and $P$-waves' contributions. 

In the present work, we will introduce the formalism of final state interaction and consider the resonance productions for the decay of $D_{s}^{+} \rightarrow K^{+} K^{-} \pi^{+}$ in Section~\ref {sec:Formalism}. Then, the calculation results and discussions are presented in Section~\ref {sec:Results}. Finally, we make a short conclusion in Section~\ref {sec:Conclusions}.

\section{Formalism}
\label{sec:Formalism}

\begin{figure}
	\begin{subfigure}{0.45\textwidth}
		\centering
		\includegraphics[width=1\linewidth,trim=150 575 200 105,clip]{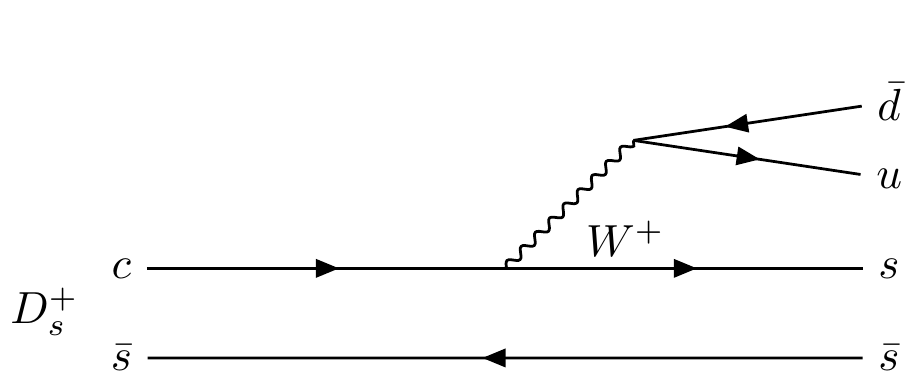} 
		\caption{\footnotesize The external $W$-emission mechanism.}
		\label{fig:Feynman1}
	\end{subfigure}
	\quad
	\quad
	\begin{subfigure}{0.45\textwidth}  
		\centering 
		\includegraphics[width=1\linewidth,trim=150 560 200 120,clip]{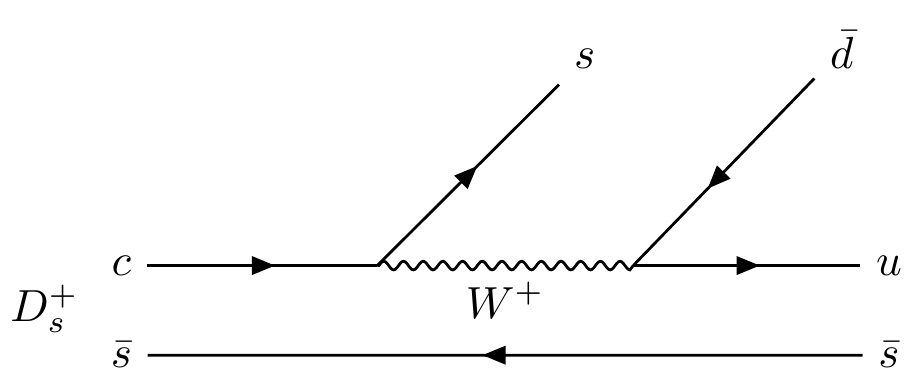} 
		\caption{\footnotesize The internal $W$-emission mechanism.}
		\label{fig:Feynman2}  
	\end{subfigure}	
	\caption{Diagrammatic representation of the $D_{s}^{+} \rightarrow K^{+} K^{-} \pi^{+}$ decay.}
	\label{fig:Feynman}
\end{figure} 

\begin{figure}
	\centering
	\includegraphics[width=0.5\linewidth,trim=150 580 170 120,clip]{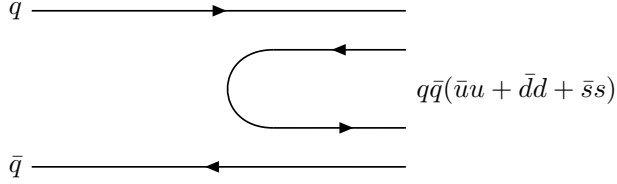}
	\caption{Schematic representation of the hadronization $q\bar{q} \rightarrow q\bar{q}(\bar{u}u+\bar{d}d+\bar{s}s)$.}
	\label{fig:Hadronization}
\end{figure}

In the present work, we investigate the decay of $D_{s}^{+} \rightarrow K^{+} K^{-} \pi^{+}$. As discussed in Refs.~\cite{Molina:2019udw,Roca:2020lyi}, the contribution for the topology diagrams of weak decay was classified by ordering their importance as follows: external $W$-emission, internal $W$-emission, $W$-exchange and annihilation, horizontal $W$-loop and vertical $W$-loop, see more details in Refs.~\cite{Chau:1982da,Chau:1987tk,Morrison:1989xq}. As implied in the experimental results~\cite{Ablikim:2020xlq}, the resonance contributions mainly involved the scalar one $S(980)$ (representing the $f_{0}(980)$ and $a_{0}(980)$ states) in $S$-wave, and the vector ones $\bar {K}^{*}(892)^{0}$ and $\phi(1020)$ in $P$-wave, whereas the contributions from the others were minor. Note that the external $W$-emission in Fig.~\ref{fig:Feynman1} can not produce the state $a_{0}(980)$ ($I=1$ ), since the $s \bar s$ pair has isospin $I=0$ as shown in the results of Ref.~\cite{Dias:2016gou}, and of course not for the $\bar {K}^{*}(892)^{0}$ with $s \bar d$, which was ignored in Ref.~\cite{Wang:2021naf}. Thus, to consider the contribution of the $\bar {K}^{*}(892)^{0}$, we need to take into account the internal $W$-emission mechanism, as shown in Fig.~\ref{fig:Feynman2}. First, let's look at the contribution of $S$-wave, as shown in Fig.~\ref{fig:Feynman1} for the external $W$-emission, where the $\bar s$ quark remains a spectator and the $c$ quark decays into an $s$ quark through the emission of a $W^{+}$ boson. Then the emission $W^{+}$ boson creates the $u \bar d$ quarks, and then eventually forms a $\pi^{+}$ meson. In order to produce the final states $K^{+}$ and $K^{-}$ alongside with the $\pi^{+}$ meson, the $s \bar s$ quark pair needs to hadronize to create a pair of pseudoscalar mesons, which can be fulfilled by the hadronization procedure through another pair of quarks generated from the vacuum, $q \bar{q} (\bar u u+\bar d d+\bar s s)$, as depicted in Fig.~\ref{fig:Hadronization}. Unlike Fig.~\ref{fig:Feynman1}, for the internal $W$-emission mechanism shown in Fig.~\ref{fig:Feynman2}, there are two possible hadronization processes. One is that the $u \bar s$ quarks form a $K^{+}$ meson and the ones $s \bar d$ undergo hadronization process. The other one is that the $s \bar d$ quarks form a $\bar K^{0}$ meson and the ones $u \bar s$ undergo the hadronization. Thus, the $D_{s}^{+}$ weak decay mechanisms in Fig.~\ref{fig:Feynman} can be formulated as follows
\begin{equation}
	\begin{aligned} 
		D^{+}_{s} & \Rightarrow V_{c s} V_{u d} (u \bar{d} \rightarrow \pi^{+}) [s\bar{s}\rightarrow s\bar{s} \cdot (\bar{u}u+\bar{d}d+\bar{s}s)]
		\\ & \Rightarrow V_{c s} V_{u d} (u \bar{d} \rightarrow \pi^{+}) [M_{33}\rightarrow (M \cdot M)_{33}],
	\end{aligned} 
	\label{eq:Dsdecay1}
\end{equation}
\begin{equation}
	\begin{aligned} 
		D^{+}_{s} & \Rightarrow V_{c s} V_{u d} \left((u \bar{s} \rightarrow K^{+}) [s\bar{d}\rightarrow s\bar{d} \cdot (\bar{u}u+\bar{d}d+\bar{s}s)]
		\right.\\ & \left.+(s \bar{d} \rightarrow \bar K^{0}) [u\bar{s}\rightarrow u\bar{s} \cdot (\bar{u}u+\bar{d}d+\bar{s}s)]	
		\right)
		\\ & \Rightarrow V_{c s} V_{u d} \left( (u \bar{s} \rightarrow K^{+}) [M_{32}\rightarrow (M \cdot M)_{32}]
		\right.\\ & \left.+(s \bar{d} \rightarrow \bar K^{0})[M_{13}\rightarrow (M \cdot M)_{13}]
			\right),
	\end{aligned} 
	\label{eq:Dsdecay2}
\end{equation}
where $V_{q_{1}q_{2}}$ is the element of the Cabibbo-Kobayashi-Maskawa (CKM) matrix from $q_{1}$ to $q_{2}$ quark, and the $q\bar{q}$ matrix element $M$ in SU(3) is defined as
\begin{equation}
	M=\left(\begin{array}{lll}{u \bar{u}} & {u \bar{d}} & {u \bar{s}} \\ {d \bar{u}} & {d \bar{d}} & {d \bar{s}} \\ {s \bar{u}} & {s \bar{d}} & {s \bar{s}}\end{array}\right).
\end{equation}

Then we can write the elements of matrix $M$ in terms of the physical mesons, which are given by
\begin{equation}
	\Phi=\left(\begin{array}{ccc}{\frac{1}{\sqrt{2}} \pi^{0}+\frac{1}{\sqrt{6}} \eta} & {\pi^{+}} & {K^{+}} \\ {\pi^{-}} & {-\frac{1}{\sqrt{2}} \pi^{0}+\frac{1}{\sqrt{6}} \eta} & {K^{0}} \\ {K^{-}} & {\bar{K}^{0}} & {-\frac{2}{\sqrt{6}} \eta}\end{array}\right),
\end{equation} 
where we take $\eta \equiv \eta_{8}$. The hadronization processes at the quark level in Eqs.~\eqref{eq:Dsdecay1} and \eqref{eq:Dsdecay2} can be accomplished to the hadron level in terms of two pseudoscalar mesons
\begin{equation}
	(M \cdot M)_{33} = (\Phi \cdot \Phi)_{33}=K^{+} K^{-}+K^{0} \bar{K}^{0}+\frac{2}{3} \eta \eta, 
	\label{eq:ss}
\end{equation} 
\begin{equation}
	(M \cdot M)_{32} = (\Phi \cdot \Phi)_{32}=K^{-} \pi^{+}- \frac{1}{\sqrt{2}}\bar{K}^{0}\pi^{0} - \frac{1}{\sqrt{6}}\bar{K}^{0}\eta,
	\label{eq:sd}
\end{equation} 
\begin{equation}
	(M \cdot M)_{13}= (\Phi \cdot \Phi)_{13}= \frac{1}{\sqrt{2}}K^{+}\pi^{0} - \frac{1}{\sqrt{6}}K^{+}\eta+K^{0}\pi^{+}.
	\label{eq:us}
\end{equation} 

Then, we get all the final states with $\pi^{+}$, $K^{+}$ or $\bar K^{0}$ produced in the hadronization processes,
\begin{equation}
	\begin{aligned}
		H^{(a)}& = V_{P} V_{c s} V_{u d}\left(K^{+} K^{-}+K^{0} \bar{K}^{0}+\frac{2}{3} \eta \eta\right) \pi^{+}
		\\ & = V_{P} V_{c s} V_{u d}\left(K^{+} K^{-}\pi^{+}+K^{0} \bar{K}^{0}\pi^{+}+\frac{2}{3} \eta \eta\pi^{+}\right) ,
	\end{aligned}
	\label{eq:Ha}
\end{equation}
\begin{equation}
	\begin{aligned}
		H^{(b)} & =\beta \times V_{P} V_{c s} V_{u d}\left[(K^{-} \pi^{+}- \frac{1}{\sqrt{2}}\bar{K}^{0}\pi^{0}
		- \frac{1}{\sqrt{6}}\bar{K}^{0}\eta) K^{+}
		\right.\\ & \left.+(\frac{1}{\sqrt{2}}K^{+}\pi^{0} - \frac{1}{\sqrt{6}}K^{+}\eta+K^{0}\pi^{+}) \bar K^{0}
		\right]
		\\ & =\beta \times V_{P} V_{c s} V_{u d}\left(K^{+}K^{-} \pi^{+}-\frac{2}{\sqrt{6}}K^{+}\bar{K}^{0}\eta 
		+K^{0}\bar K^{0}\pi^{+}
		\right),
	\end{aligned}
	\label{eq:Hb}
\end{equation}
where $H^{(a)}$ and $H^{(b)}$ are the contributions of Fig.~\ref{fig:Feynman1} and Fig.~\ref{fig:Feynman2}, respectively. The $V_P$ is the production vertex factor of weak decay process in Fig. \ref{fig:Feynman1}, which contains all the dynamical factors in the weak interaction and is an unknown parameter in our formalism, see the discussion later. In our calculation, we take $V_P$ as a constant~\cite{Wang:2020pem,Ahmed:2020qkv}. For the internal $W$-emission in Fig.~\ref{fig:Feynman2}, the production vertex factor should be different, and thus one more coefficient $\beta$ in $H^{(b)}$ is introduced for the relative weight of the internal $W$-emission mechanism with respect to the external one~\cite{Dai:2018nmw,Wang:2020pem}. In Fig.~\ref{fig:Feynman1}, the $u$ and $\bar{d}$ quarks from the external $W$-emission are constrained to form the color singlet $\pi^{+}$. Whereas, in Fig.~\ref{fig:Feynman2}, the $u$ and $\bar{d}$ quarks from internal $W$-emission have the fixed colors. Thus, the absolute value of $\beta$ should be less than $1$. Then, we obtain the total contributions
\begin{equation}
	\begin{aligned}
		H & = H^{(a)}+H^{(b)}
		\\ & =  V_{P} V_{c s} V_{u d} \left[(1+\beta)K^{+} K^{-}\pi^{+}+(1+\beta)K^{0}\bar{K}^{0}\pi^{+}+\frac{2}{3} \eta \eta\pi^{+}
		 - \frac{2\beta}{\sqrt{6}}K^{+}\bar{K}^{0}\eta \right].
	\end{aligned}
	\label{eq:H}
\end{equation}

\begin{figure}
	\begin{subfigure}{1\textwidth}
		\centering
		\includegraphics[width=0.5\linewidth,trim=120 540 190 80,clip]{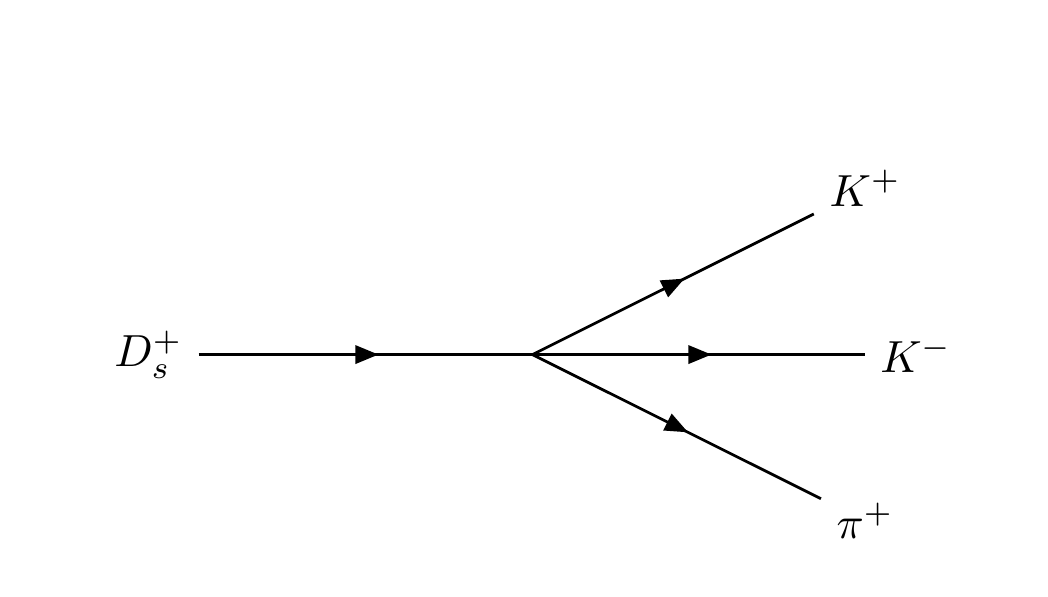} 
		\caption{\footnotesize Tree-level production.}
		\label{fig:Scatter1}
	\end{subfigure}
	\quad
	\quad
	\begin{subfigure}{1\textwidth}  
		\centering 
		\includegraphics[width=0.45\linewidth,trim=150 530 180 130,clip]{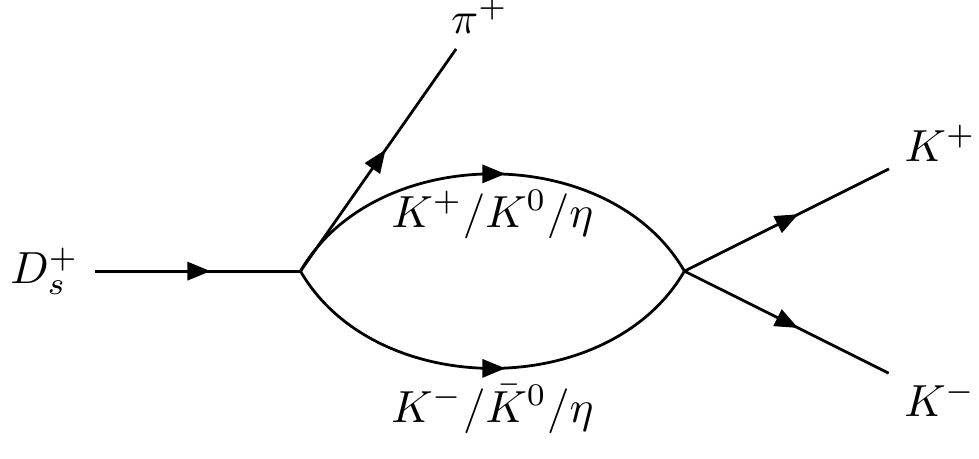} 
		\caption{\footnotesize Rescattering of $K^{+} K^{-}$, $K^{0} \bar{K}^{0}$ and $\eta \eta$.}
		\label{fig:Scatter2}  
	\end{subfigure}	
	\quad
	\quad
	\begin{subfigure}{1\textwidth}  
		\centering 
		\includegraphics[width=0.45\linewidth,trim=150 540 180 130,clip]{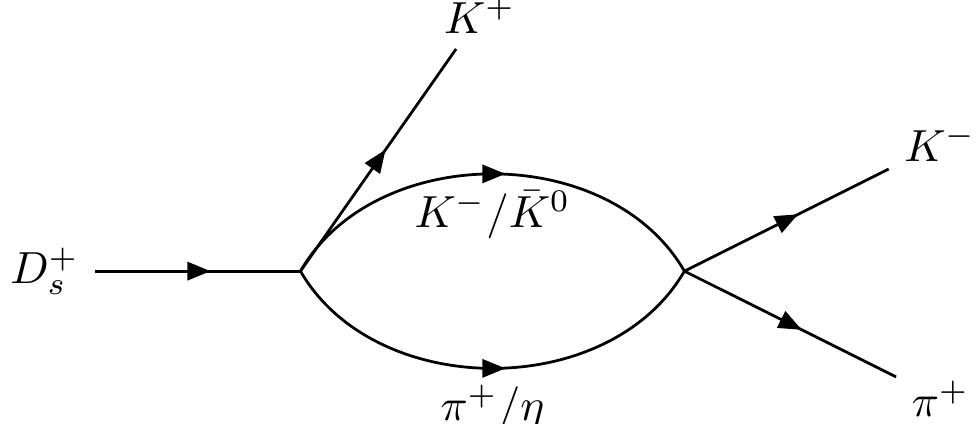} 
		\caption{\footnotesize Rescattering of $K^{-}\pi^{+}$ and $\bar{K}^{0}\eta$.}
		\label{fig:Scatter3}  
	\end{subfigure}	
	\caption{Diagrammatic representation of the $D_{s}^{+} \rightarrow K^{+} K^{-} \pi^{+}$ decay.}
	\label{fig:Scatter}
\end{figure} 

One can see in Eq.~\eqref{eq:H} that the final states $K^{+} K^{-} \pi^{+}$ are produced directly in the hadronization processes. For the other ones, taking into account the final state interactions, we can also generate the final states $K^{+} K^{-} \pi^{+}$ by the rescattering procedure, as depicted in Fig.~\ref{fig:Scatter}, see more discussions in Ref.~\cite{Miyahara:2015cja}. Thus, the amplitudes for Fig.~\ref{fig:Scatter} with these final state productions at the tree level and rescattering processes can be written as
\begin{equation}
	\begin{aligned} 
		t(s_{12},s_{23})=\mathcal D & \left [ (1+\beta)+(1+\beta)G_{K^{+} K^{-}}(s_{12}) T_{K^{+} K^{-} \rightarrow K^{+} K^{-}}(s_{12})
		\right.\\ & \left.+(1+\beta)G_{K^{0} \bar{K}^{0}}(s_{12})  T_{K^{0} \bar{K}^{0} \rightarrow K^{+} K^{-}}(s_{12})
		\right.\\ & \left.+\frac{2}{3} G_{\eta \eta}(s_{12})  T_{\eta \eta \rightarrow K^{+} K^{-}}(s_{12})
		\right.\\ & \left.+(1+\beta)G_{K^{-}\pi^{+}}(s_{23})  T_{K^{-}\pi^{+}\rightarrow K^{-}\pi^{+}}(s_{23}) 
		\right.\\ & \left.-\frac{2\beta}{\sqrt{6}} G_{\bar{K}^{0}\eta}(s_{23})  T_{\bar{K}^{0}\eta \rightarrow K^{-}\pi^{+}}(s_{23})
		\right],
	\end{aligned}
	\label{eq:amplitudes}
\end{equation}
where the factors $V_{P}$, $V_{c s}$ and $V_{u d}$ have been absorbed in the constant $\mathcal D$, which is also included the normalization factor when we fit the invariant mass distributions later. And the energy of two-body system is defined as $s_{ij}=(p_{i}+p_{j})^2$, with $p_{i}$ and $p_{j}$ the four-momenta of the two particles, where the indices $i,\ j=1,\ 2,\ 3$ denote the three final states of $K^+$, $K^-$, $\pi^{+}$, respectively. Besides, $G_{ij}$ is the loop function of two mesons propagators, which will be discussed later. It is worth to mention that, in Eq.~\eqref{eq:amplitudes}, in principle there is a factor of $2$ in the term related with the identical particles $\eta\eta$ because of the two possibilities in the operators of Eq.~\eqref{eq:H} to create them, which has been cancelled with the factor of $1/2$ in their propagators within our normalization scheme, see more discussions in Ref.~\cite{Liang:2015qva}. Note that in Eq.~\eqref{eq:amplitudes}, the channels $K^{+} K^{-}$ and $K^{0} \bar{K}^{0}$ in $S$-wave can be decomposition with isospins $I=0$ and $I=1$, written
\begin{equation}
T_{K^{+} K^{-}\to K^{+} K^{-}}=\frac{1}{2} \left(T^{I=0}_{K\bar{K} \to K\bar{K}} + T^{I=1}_{K\bar{K} \to K\bar{K}}\right), 
\label{eq:tk1k2}
\end{equation}
\begin{equation}
T_{K^{0} \bar{K}^{0}\to K^{+} K^{-}}=\frac{1}{2} \left(T^{I=0}_{K\bar{K} \to K\bar{K}} - T^{I=1}_{K\bar{K} \to K\bar{K}}\right),
\label{eq:tk0kb0}
\end{equation}
where we have used the convention of the physical states $\left|\pi^{+}\rangle=-\right| 1,1\rangle$ and $\left|K^{+}\rangle=-\right| 1/2,1/2\rangle$ for the isospin basis~\cite{Oller:1997ti}. But, from the second and third parts of Eq.~\eqref{eq:amplitudes} and Eqs.~\eqref{eq:tk1k2}, \eqref{eq:tk0kb0}, one can see that the summation finally only with isospin $I = 0$ contributes to the $K^{+}K^{-}$ final state interaction for the $D_{s}^{+} \rightarrow K^{+} K^{-} \pi^{+}$ decay, and thus there is no $I = 1$ component, which is consistent with the findings of Refs.~\cite{delAmoSanchez:2010yp,Wang:2021naf}. Therefore, there should be only the resonance $f_0(980)$ contribution in the $K^{+}K^{-}$ invariant mass distribution, and without the  one of $a_0(980)$. Indeed, as found in Ref.~\cite{Xiao:2019lrj}, the $f_0(980)-a_0(980)$ mixing would not appear in the scattering amplitudes of $T_{K^{+} K^{-}\to K^{+} K^{-}}$ and $T_{K^{0} \bar{K}^{0}\to K^{+} K^{-}}$, see Eqs.~\eqref{eq:tk1k2} and \eqref{eq:tk0kb0}, which can also be seen in our results later. Moreover, the ingredients $\pi^{+} K^{-}$ in $S$-wave can be in $I = 1/2$ and $I = 3/2$. But as found in Ref.~\cite{Toledo:2020zxj}, the $I = 1/2$ is dominant, and thus we omit the contribution of $I = 3/2$. 

Furthermore, the two-body scattering amplitudes in Eq.~\eqref{eq:amplitudes}, such as $T_{K^{+} K^{-}\to K^{+} K^{-}}$, can be evaluated by the coupled channel Bethe-Salpeter equation of the ChUA,
\begin{equation}
	\begin{aligned} 
		T = [1-VG]^{-1}V, 
	\end{aligned}
	\label{eq:BSE}
\end{equation}
where the matrix $V$ is constructed by the scattering potentials for each coupled channel. In the ChUA, the interaction potentials can be calculated from the chiral Lagrangians. 
Since there is no $I=1$ contribution, there are five channels coupled to $K^{+} K^{-}$, $\pi^{+}\pi^{-}$, $\pi^{0}\pi^{0}$, $K^{+} K^{-}$, $K^{0} \bar{K}^{0}$, and $\eta\eta$, which are denoted as 1 to 5 accordingly and given by (after applying the $S$-wave projection)~\cite{Ahmed:2020qkv},
\begin{equation}
	\begin{aligned}
		&V_{11}=-\frac{1}{2 f^{2}} s, \quad V_{12}=-\frac{1}{\sqrt{2} f^{2}}\left(s-m_{\pi}^{2}\right), \quad V_{13}=-\frac{1}{4 f^{2}} s ,\\
		&V_{14}=-\frac{1}{4 f^{2}} s, \quad V_{15}=-\frac{1}{3 \sqrt{2} f^{2}} m_{\pi}^{2}, \quad V_{22}=-\frac{1}{2 f^{2}} m_{\pi}^{2} ,\\
		&V_{23}=-\frac{1}{4 \sqrt{2} f^{2}} s, \quad V_{24}=-\frac{1}{4 \sqrt{2} f^{2}} s, \quad V_{25}=-\frac{1}{6 f^{2}} m_{\pi}^{2} ,\\
		&V_{33}=-\frac{1}{2 f^{2}} s, \quad V_{34}=-\frac{1}{4 f^{2}} s ,\\
		&V_{35}=-\frac{1}{12 \sqrt{2} f^{2}}\left(9 s-6 m_{\eta}^{2}-2 m_{\pi}^{2}\right), \quad V_{44}=-\frac{1}{2 f^{2}} s ,\\
		&V_{45}=-\frac{1}{12 \sqrt{2} f^{2}}\left(9 s-6 m_{\eta}^{2}-2 m_{\pi}^{2}\right) ,\\
		&V_{55}=-\frac{1}{18 f^{2}}\left(16 m_{K}^{2}-7 m_{\pi}^{2}\right),
	\end{aligned}
\end{equation}  
where $f$ is the pion decay constant, taken as $f=0.093$ GeV \cite{Oller:1997ti}, and $m_P$ the corresponding mass of pseudoscalar meson ($P$). For the $I=1/2$ sector, there are three coupled channels, $K^{-}\pi^{+}$, $\bar{K}^{0}\pi^{0}$, and $\bar{K}^{0}\eta$, which are specified as 1 to 3 channels, respectively, and given by~\cite{Toledo:2020zxj}
\footnote{Note that in Ref. \cite{Toledo:2020zxj} the coupled channels are $K^{+}\pi^{-}$, ${K}^{0}\pi^{0}$, and ${K}^{0}\eta$, which are the same for our case due to the charge symmetry.},
\begin{equation}
	\begin{aligned}
		&V_{11}=\frac{-1}{6 f^{2}}\left(\frac{3}{2} s-\frac{3}{2 s}\left(m_{\pi}^{2}-m_{K}^{2}\right)^{2}\right), \\
		&V_{12}=\frac{1}{2 \sqrt{2} f^{2}}\left(\frac{3}{2} s-m_{\pi}^{2}-m_{K}^{2}-\frac{\left(m_{\pi}^{2}-m_{K}^{2}\right)^{2}}{2 s}\right), \\
		&V_{22}=\frac{-1}{4 f^{2}}\left(-\frac{s}{2}+m_{\pi}^{2}+m_{K}^{2}-\frac{\left(m_{\pi}^{2}-m_{K}^{2}\right)^{2}}{2 s}\right), \\
		&V_{13}=\frac{1}{2 \sqrt{6} f^{2}}\left(\frac{3}{2} s-\frac{7}{6} m_{\pi}^{2}-\frac{1}{2} m_{\eta}^{2}-\frac{1}{3} m_{K}^{2}+\frac{3}{2 s}\left(m_{\pi}^{2}-m_{K}^{2}\right)\left(m_{\eta}^{2}-m_{K}^{2}\right)\right), \\
		&V_{23}=-\frac{1}{4 \sqrt{3} f^{2}}\left(\frac{3}{2} s-\frac{7}{6} m_{\pi}^{2}-\frac{1}{2} m_{\eta}^{2}-\frac{1}{3} m_{K}^{2}+\frac{3}{2 s}\left(m_{\pi}^{2}-m_{K}^{2}\right)\left(m_{\eta}^{2}-m_{K}^{2}\right)\right), \\
		&V_{33}=-\frac{1}{4 f^{2}}\left(-\frac{3}{2} s-\frac{2}{3} m_{\pi}^{2}+m_{\eta}^{2}+3 m_{K}^{2}-\frac{3}{2 s}\left(m_{\eta}^{2}-m_{K}^{2}\right)^{2}\right).
	\end{aligned}
\end{equation}

On the other hand, the diagonal matrix $G$ is made up by the loop functions, of which the element is two intermediate meson propagators, given by
\begin{equation}
		G _ {kk} ( s ) = i \int \frac { d ^ { 4 } q } { ( 2 \pi ) ^ { 4 } } \frac { 1 } { q ^ { 2 } - m _ { 1 } ^ { 2 } + i \varepsilon } \frac { 1 } { \left( p _ { 1 } + p _ { 2 } - q \right) ^ { 2 } - m _ { 2 } ^ { 2 } + i \varepsilon }  \text{ ,}
	\label{eq:propagators}
\end{equation}
where $p_{1}$ and $p_{2}$ are the four-momenta of the two initial particles in the $k$-th channel, $m_{1}$ and $m_{2}$ are the masses of two intermediate particles, and $s=(p_{1}+p_{2})^2$. Since the integral of this equation is logarithmically divergent, we take the explicit form of dimensional regularization method~\cite{Oller:2000fj},
\begin{equation}
	\begin{aligned}
		G_{kk}(s)=& \frac{1}{16 \pi^{2}}\left\{a_{\mu}+\ln \frac{m_{1}^{2}}{\mu^{2}}+\frac{m_{2}^{2}-m_{1}^{2}+s}{2 s} \ln \frac{m_{2}^{2}}{m_{1}^{2}}\right.\\
		&+\frac{q_{cm\,k}(s)}{\sqrt{s}}\left[\ln \left(s-\left(m_{2}^{2}-m_{1}^{2}\right)+2 q_{cm\,k}(s) \sqrt{s}\right)\right.\\
		&+\ln \left(s+\left(m_{2}^{2}-m_{1}^{2}\right)+2 q_{cm\,k}(s) \sqrt{s}\right) \\
		&-\ln \left(-s-\left(m_{2}^{2}-m_{1}^{2}\right)+2 q_{cm\,k}(s) \sqrt{s}\right) \\
		&\left.\left.-\ln \left(-s+\left(m_{2}^{2}-m_{1}^{2}\right)+2 q_{cm\,k}(s) \sqrt{s}\right)\right]\right\},
	\end{aligned}
	\label{eq:DR}
\end{equation}
where $\mu$ is the regularization scale, chosen as 0.6 GeV from Refs.~\cite{Duan:2020vye,Ahmed:2020qkv}, and $a_{\mu}$ the subtraction constant. Following the method of Ref.~\cite{Oset:2001cn}, we will determine the values of $a_{\mu}$ for different channels, see the discussion later. Besides, $q_{cm\,k}(s)$ is the three momentum of the particle in the center-of-mass frame, given by 
\begin{equation}
		q_{cm\,k}(s)=\frac{\lambda^{1 / 2}\left(s, m_{1}^{2}, m_{2}^{2}\right)}{2 \sqrt{s}},
	\label{eq:qcmi}
\end{equation}
with the usual Källen triangle function $\lambda(a, b, c)=a^{2}+b^{2}+c^{2}-2(a b+a c+b c)$.

Note that, in Fig.~\ref{fig:Feynman1}, the $s\bar{s}$ quarks are not only in $S$-wave to form $K^{+}K^{-}$ through hadronization
\footnote{One should keep in mind that the $f_0(980)$ resonance is generated in the final state interaction.}, 
but also in $P$-wave to produce the vector meson $\phi(1020)$, which decays into $K^{+}K^{-}$ finally. We consider a full relativistic amplitude for the decay $D_{s}^{+} \rightarrow \phi(1020) \pi^{+} \rightarrow K^{+} K^{-} \pi^{+}$ as done in Refs.~\cite{Toledo:2020zxj,Roca:2020lyi},
\begin{equation}
		\mathcal M_{\phi}(s_{12},s_{23})={\mathcal D_{\phi}}e^{i\alpha_{\phi}} \frac{s_{23}-s_{13}}{s_{12}-m_{\phi}^{2}+i m_{\phi} \Gamma_{\phi}},
	\label{eq:amplitudephi}
\end{equation}
where $\mathcal D_{\phi}$ is a normalization constant and $\alpha_{\phi}$ a phase, which will be fitted by the experimental data later. Besides, $\Gamma_{\phi}$ is the total width of the $\phi(1020)$, taking as $\Gamma_{\phi}=4.25$ MeV. One thing should be mentioned that the $s_{ij}$ are not independent totally and fulfill the constraint condition,
\begin{equation}
		s_{12}+s_{23}+s_{13}=m_{D_{s}^{+}}^{2}+m_{K}^{2}+m_{K}^{2}+m_{\pi}^{2},
	\label{eq:sij}
\end{equation}
which means that only two of $s_{ij}$ variables are independent.

Analogously, in Fig. \ref{fig:Feynman2}, the $s\bar{d}$ quarks can form the $\bar {K}^{*}(892)^{0}$ meson in $P$-wave, and then $\bar {K}^{*}(892)^{0}$ decays into $K^{-}\pi^{+}$. The full relativistic amplitude for the process $D_{s}^{+} \rightarrow K^{+} \bar {K}^{*}(892)^{0} \rightarrow K^{+} K^{-} \pi^{+}$ is given by~\cite{Toledo:2020zxj}
\begin{equation}
	\begin{aligned}
		\mathcal M_{\bar{K}^{*}}(s_{12},s_{23})= \frac{\mathcal D_{\bar{K}^{*}}e^{i\alpha_{\bar{K}^{*}}}} {s_{23}-m_{\bar{K}^{*}}^{2}+i m_{\bar{K}^{*}} \Gamma_{\bar{K}^{*}}}\left[(m_{K}^{2}-m_{\pi}^{2})\frac{m_{D_{s}^{+}}^{2}-m_{K}^{2}}{m_{\bar{K}^{*}}^{2}}+s_{13}-s_{12}\right],
	\end{aligned}
	\label{eq:amplitudeKstar}
\end{equation}
where $\mathcal D_{\bar {K}^{*}}$ is a constant, $\alpha_{\bar {K}^{*}}$ a phase, $\Gamma_{\bar {K}^{*}}$ the total width of the $\bar {K}^{*}(892)^{0}$, taking as $\Gamma_{\bar {K}^{*}}=50.80$ MeV. Then the total amplitude with the contributions of $S$- and $P$-waves is obtained as 
\begin{equation}
		t^{\prime}(s_{12},s_{23})=t(s_{12},s_{23})+\mathcal{M}_{\phi}(s_{12},s_{23})+\mathcal{M}_{\bar{K}^{*}}(s_{12},s_{23}).
	\label{eq:tprime}
\end{equation}

Finally, we get the double differential width distribution of three-body decay~\cite{Zyla:2020pdg},
\begin{equation}
		\frac{d^{2} \Gamma}{d s_{12}d s_{23}}=\frac{1}{(2 \pi)^{3}} \frac{1}{32 m_{D_{s}^{+}}^{3}}\left|t^{\prime}(s_{12},s_{23})\right|^{2}.
	\label{eq:dGamma}
\end{equation}
Thus, the single invariant mass distributions $d \Gamma/ds_{12}$ and $d \Gamma/ds_{23}$ can be obtained by integrating the other invariant mass variable in Eq.~\eqref{eq:dGamma}. Furthermore, one can obtain $d \Gamma/ds_{13}$ through Eq.~\eqref{eq:sij}. For integrating $s_{23}$, the limits of integration are given in Particle Data Group (PDG)~\cite{Zyla:2020pdg}, written
\begin{equation}
		\left(s_{23}\right)_{\max}=\left(E_{2}^{*}+E_{3}^{*}\right)^{2}-\left(\sqrt{E_{2}^{*2}-m_{2}^{2}}-\sqrt{E_{3}^{*2}-m_{3}^{2}}\right)^{2}, 
\end{equation}
\begin{equation}
		\left(s_{23}\right)_{\min } =\left(E_{2}^{*}+E_{3}^{*}\right)^{2}-\left(\sqrt{E_{2}^{* 2}-m_{2}^{2}}+\sqrt{E_{3}^{* 2}-m_{3}^{2}}\right)^{2},
\end{equation}
where
\begin{equation}
		E_{2}^{*}=\frac{s_{12}-m_{1}^{2}+m_{2}^{2}}{2 \sqrt{s_{12}}}, 
\end{equation}
\begin{equation}
		E_{3}^{*}=\frac{m_{D_{s}^{+}}^{2}-s_{12}-m_{3}^{2}}{2 \sqrt{s_{12}}}.
\end{equation}
And for integrating $s_{12}$, the limits of integration are given by 
\begin{equation}
		\left(s_{12}\right)_{\max }=\left(E_{2}^{*^{\prime}}+E_{1}^{*^{\prime}}\right)^{2}-\left(\sqrt{E_{2}^{*^{\prime\; 2}}-m_{2}^{2}}-\sqrt{E_{1}^{*^{\prime\; 2}}-m_{1}^{2}}\right)^{2}, 
\end{equation}
\begin{equation}
		\left(s_{12}\right)_{\min }=\left(E_{2}^{*^{\prime}}+E_{1}^{*^{\prime}}\right)^{2}-\left(\sqrt{E_{2}^{*^{\prime\; 2}}-m_{2}^{2}}+\sqrt{E_{1}^{*^{\prime\; 2}}-m_{1}^{2}}\right)^{2},
\end{equation}
where
\begin{equation}
		E_{2}^{*^{\prime}}=\frac {s_{23}-m_{3}^{2}+m_{2}^{2}}{2 \sqrt {s_{23}}}, 
\end{equation}
\begin{equation}
		E_{1}^{*^{\prime}}=\frac {m_{D_{s}^{+}}^{2}-s_{23}-m_{1}^{2}} {2 \sqrt {s_{23}}}.
\end{equation}

Note that in the ChUA, we can only make reliable predictions up to $1.1\sim 1.2$ GeV in the coupled channel interaction. The limits of the integral variable of Eq.~\eqref{eq:dGamma} for the invariant masses are higher than $1.2$ GeV. Of course, we are more interested in the resonance region of the invariant masses below $1.1$ GeV, where the region above $1.1$ GeV has little impact, see the results of Ref.~\cite{Wang:2021naf}. Thus, as done in Refs.~\cite{Debastiani:2016ayp,Toledo:2020zxj}, we smoothly extrapolate $G\left(s\right)T\left(s\right)$ above the energy cut $\sqrt{s} \geq \sqrt{s_{cut}}=1.1$ GeV using
\begin{equation}
	G\left(s\right)T\left(s\right)=G\left(s_{cut}\right)T\left(s_{cut}\right){e}^{-\alpha\left(\sqrt{s}-\sqrt{s_{cut}}\right)}, \text { for } \sqrt{s}>\sqrt{s_{cut}} ,
	\label{eq:GT}
\end{equation}
where $G$ is the loop function of two meson propagators in Eq.~\eqref{eq:DR}, and $T$ the amplitude obtained by Eq.~\eqref{eq:BSE}. Besides, $\alpha$ is a smoothing extrapolation parameter, of which the value will be discussed in the next section.

\section{Results}
\label{sec:Results}

In our calculations, we first determine the values of subtraction constants $a_{\mu}$ in the loop functions. Following the method of Ref.~\cite{Oset:2001cn}, one can make the loop functions to have the same value at the threshold of each coupled channel using the dimensional regularization and the cut off formulae, and then determine the values of the subtraction constants $a_{\mu}$ for each coupled channel. For $I = 0$ sector, taking the cutoff $q_{max} = 0.6$ GeV~\cite{Duan:2020vye,Ahmed:2020qkv}, we obtain
\begin{equation}
	\begin{aligned}
		a_{\pi^{+} \pi^{-}}=-1.30,\ a_{\pi^{0} \pi^{0}}=-1.29,\ a_{K^{+} K^{-}}=-1.63,\ a_{K^{0} \bar K^{0}}=-1.63,\ a_{\eta \eta}=-1.68.
	\end{aligned}
\end{equation}
For $I = 1/2$ sector, based on the cutoff obtained in Ref.~\cite{Guo:2005wp}, which is also $q_{max} = 0.6$ GeV, we have
\begin{equation}
	\begin{aligned}
		a_{\pi^{+} K^{-}}=-1.57,\ a_{\pi^{0} \bar K^{0}}=-1.57,\ a_{\eta \bar K^{0}}=-1.66.
	\end{aligned}
\end{equation}
Note that, as discussed in Ref.~\cite{Ahmed:2020qkv}, in order to investigate the properties of the resonances, we do not treat $a_{\mu}$ as free parameter in our fits. Thus, the resonances $f_0(980)$ and $K_0^*(700)$ (also called $\kappa$) are naturally generated in the coupled channel interactions of the ChUA.

\begin{table}[htbp]
	\centering
	\caption{Values of the parameters from the fit.}
	\resizebox{1.0\textwidth}{!}
	{\begin{tabular}{cccccccc}
			\hline \hline 
			Parameters &\quad $\mathcal D$&\quad $\beta$&\quad $\alpha$&\quad $\mathcal D_{\phi}$&\quad $\alpha_{\phi}$&\quad $\mathcal D_{\bar{K}^{*}}$&\quad $\alpha_{\bar{K}^{*}}$   \\
			\hline 
			Fit \uppercase\expandafter{\romannumeral1} &\quad $6635.79$&\quad $0.16$&\quad $19.62$&\quad $1201.86$&\quad $0.00$(fixed)&\quad $825.97$&\quad $0.15$ \\
			
			Fit \uppercase\expandafter{\romannumeral2} &\quad $3151.32$&\quad $-0.90$&\quad $7.13$&\quad $158.32$&\quad $0.39$&\quad $148.56$&\quad $0.05$ \\
			\hline\hline  
	\end{tabular}}
	\label{tab:Parameters}
\end{table}

\begin{figure}[htbp]
	\begin{subfigure}{0.47\textwidth}
		\centering
		\includegraphics[width=1\linewidth]{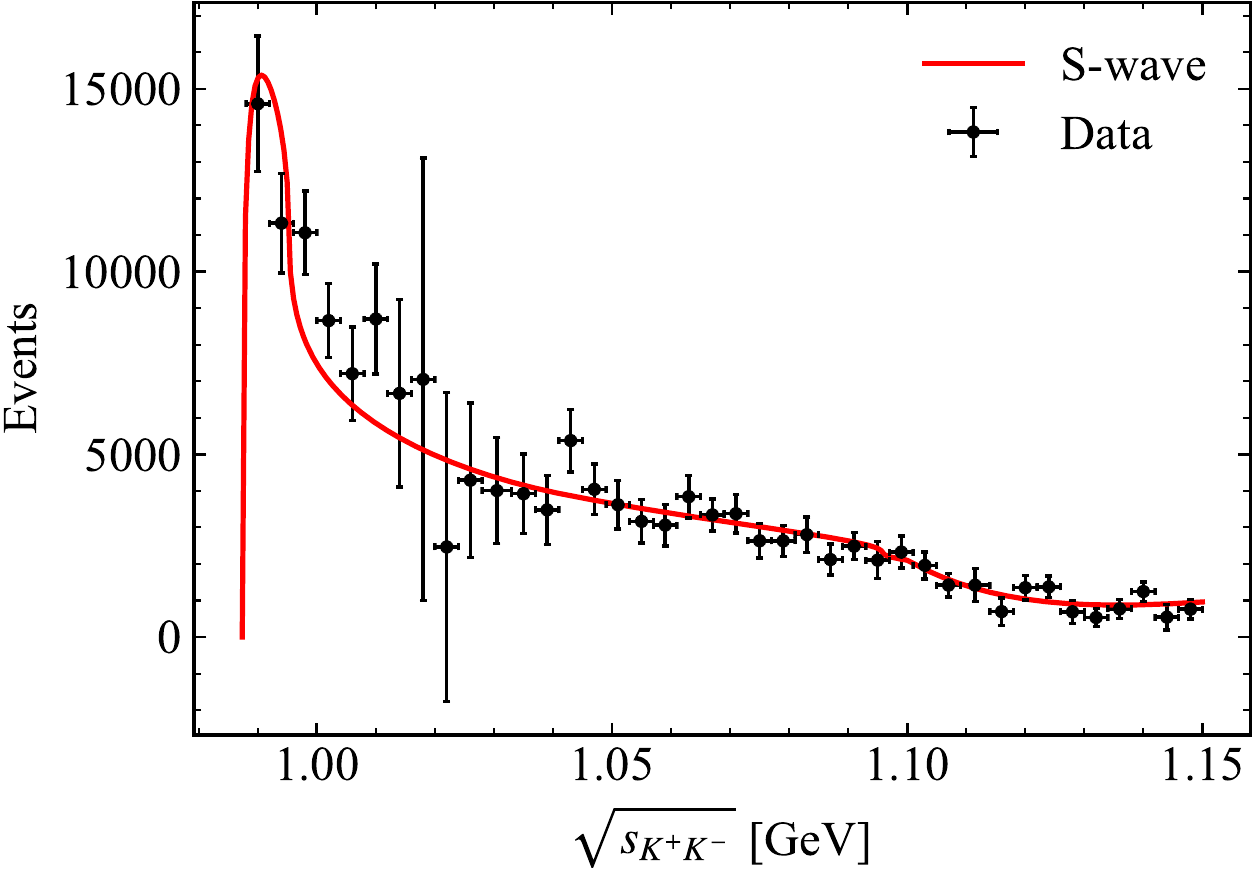} 
		\caption{\footnotesize Results for $S$-wave with $\chi^{2}/dof.=37.51/(40-3)=1.01$.}
		\label{fig:SMinv12}
	\end{subfigure}
	\quad
	\begin{subfigure}{0.47\textwidth}  
		\centering 
		\includegraphics[width=1\linewidth]{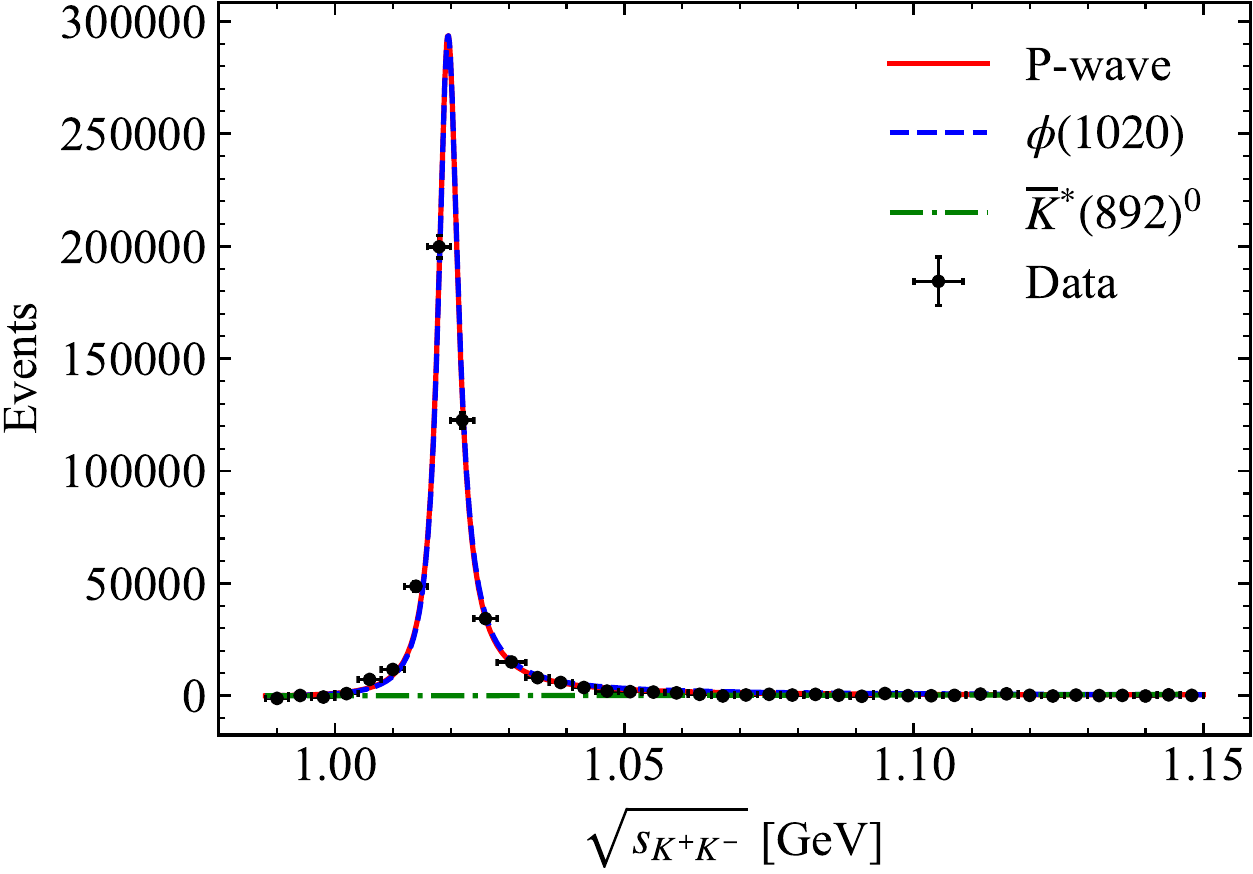} 
		\caption{\footnotesize Results for $P$-wave with $\chi^{2}/dof.=206.61/(40-3)=5.58$.}
		\label{fig:PMinv12}  
	\end{subfigure}	
	\caption{$K^{+}K^{-}$ invariant mass distributions in $S$-wave and $P$-wave of the $D_{s}^{+} \to K^{+} K^{-} \pi^{+}$ decay (with the reduced chi-square). Data are taken from Ref.~\cite{Ablikim:2020xlq}.}
	\label{fig:Minv12}
\end{figure} 

Therefore, we have seven parameters in the fit of the experimental data, $\mathcal D$, $\beta$ and $\alpha$ in $S$-wave, ${\mathcal D_{\phi}}$, $\alpha_{\phi}$, ${\mathcal D_{\bar{K}^{*}}}$ and $\alpha_{\bar{K}^{*}}$ in $P$-wave. But, some of these parameters are uncorrelated, since there is no interference effect between $S$- and $P$-waves. First, we fit the $K^{+} K^{-}$ invariant mass distributions in $S$- and $P$-waves separately (the parameters for the other one are set as zero), which are shown in Fig.~\ref{fig:Minv12}. From Fig.~\ref{fig:SMinv12}, we can obtain the parameters of $S$-wave, and the ones of $P$-wave from Fig.~\ref{fig:PMinv12}. The fitting parameters are shown in Fit~\uppercase\expandafter{\romannumeral1} results of Table~\ref{tab:Parameters}, where the phase of $\phi(1020)$, $\alpha_{\phi}$, is fixed as $0$ in the $P$-wave fitting. Besides, we can see that the fitting value of $\beta$ is $0.16$, which is less than $1$ as discussed before. Note that, the uncertainties for the parameters of $S$-wave are quite large due to its small contribution as found later. One can see from Fig.~\ref{fig:Minv12} that the fitting results for the $K^{+} K^{-}$ invariant mass distributions are in good agreement with the experimental data both in $S$- and $P$-waves. Our results of Fig.~\ref{fig:SMinv12} for $S$-wave are consistent with those obtained in Ref. \cite{Wang:2021naf}. Note that we use the dimensional regularization method to solve singular integration of the loop function, see Eq.~\eqref{eq:DR}, which is different with the one of cutoff method used in Ref.~\cite{Wang:2021naf}. On the other hand, we use the double differential width distribution of three-body decay in Eq.~\eqref{eq:dGamma}, whereas, only the two-body $K^+ K^-$ invariant mass distribution was concerned in Ref.~\cite{Wang:2021naf}. Thus, our results include the contribution of $K^{-}\pi^{+}$ channel, although it is very small, see the results later. As shown in Fig.~\ref{fig:PMinv12}, indeed the main contribution is the $\phi$ in the $P$-wave $K^+ K^-$ invariant mass distribution. 

\begin{figure}[htbp]
	\centering
	\includegraphics[width=0.6\linewidth]{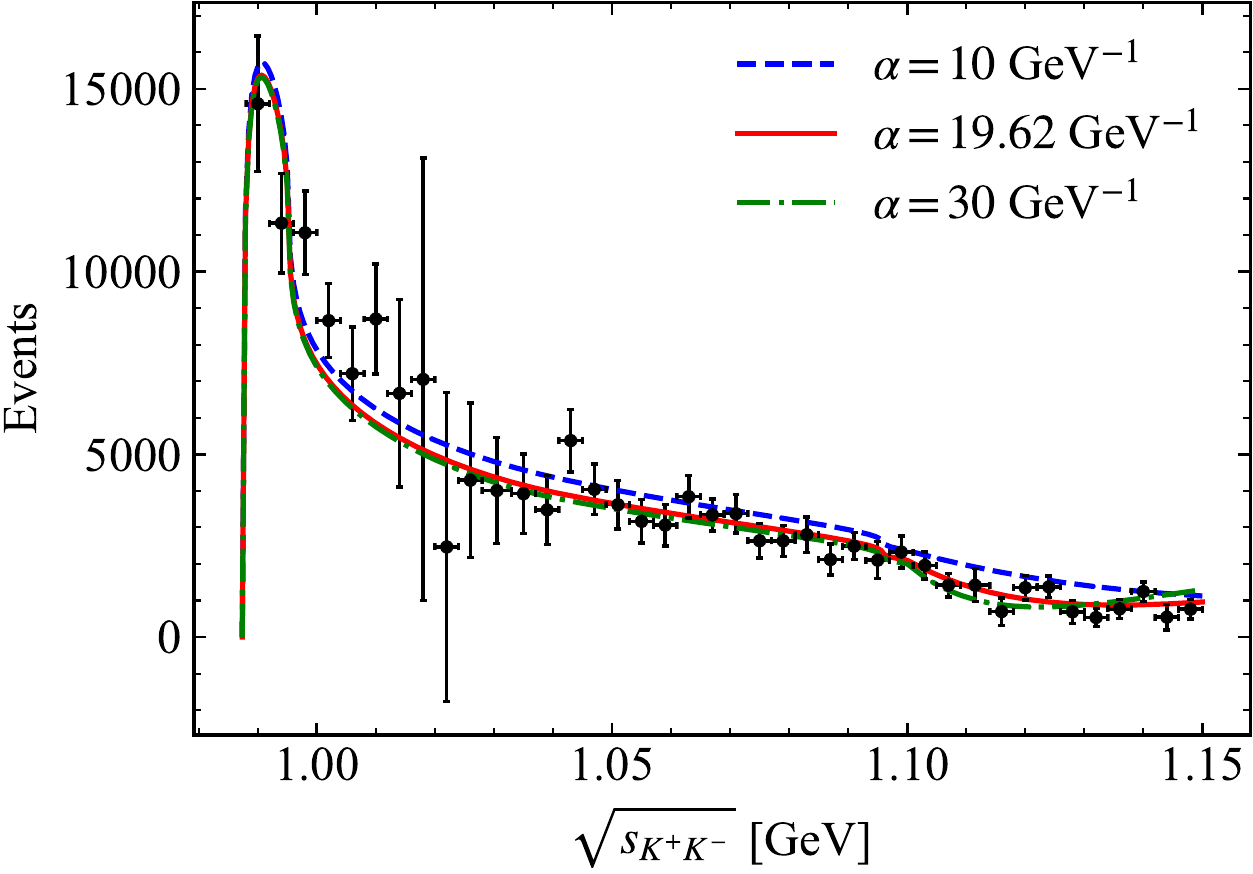} 
	\caption{Effects of the different extrapolation parameter $\alpha$ on the $S$-wave $K^{+}K^{-}$ invariant mass distribution.}
	\label{fig:SMinv12alpha}
\end{figure} 

In our fitting, the value of $\alpha$ in Eq.~\eqref{eq:GT} is $19.62$ GeV$^{-1}$ (see Table~\ref{tab:Parameters}), which means that $G\left(s\right)T\left(s\right)$ is reduced by seven times at the position of $\sqrt{s_{cut}}+0.1$ GeV. To check the effects of the smooth extrapolation of the amplitudes, we also take $\alpha=10$ GeV$^{-1}$ and $30$ GeV$^{-1}$, which reduce $G\left(s\right)T\left(s\right)$ by a factor about 3 and 19, respectively, at $\sqrt{s_{cut}}+0.1$ GeV. All the other parameters are taken the same values as the ones in Fit~\uppercase\expandafter{\romannumeral1} of Table~\ref{tab:Parameters}. Our results are shown in Fig.~\ref{fig:SMinv12alpha}, where one can see that the results barely change in the $S$-wave $K^{+} K^{-}$ invariant mass distribution with different extrapolation factors (different value of $\alpha$). This indicates that the influence of $\alpha$ on our fitting results is trivial, since it only affects the data above the cut at 1.1 GeV.

\begin{figure}[htbp]
	\begin{subfigure}{0.47\textwidth}
		\centering
		\includegraphics[width=1\linewidth]{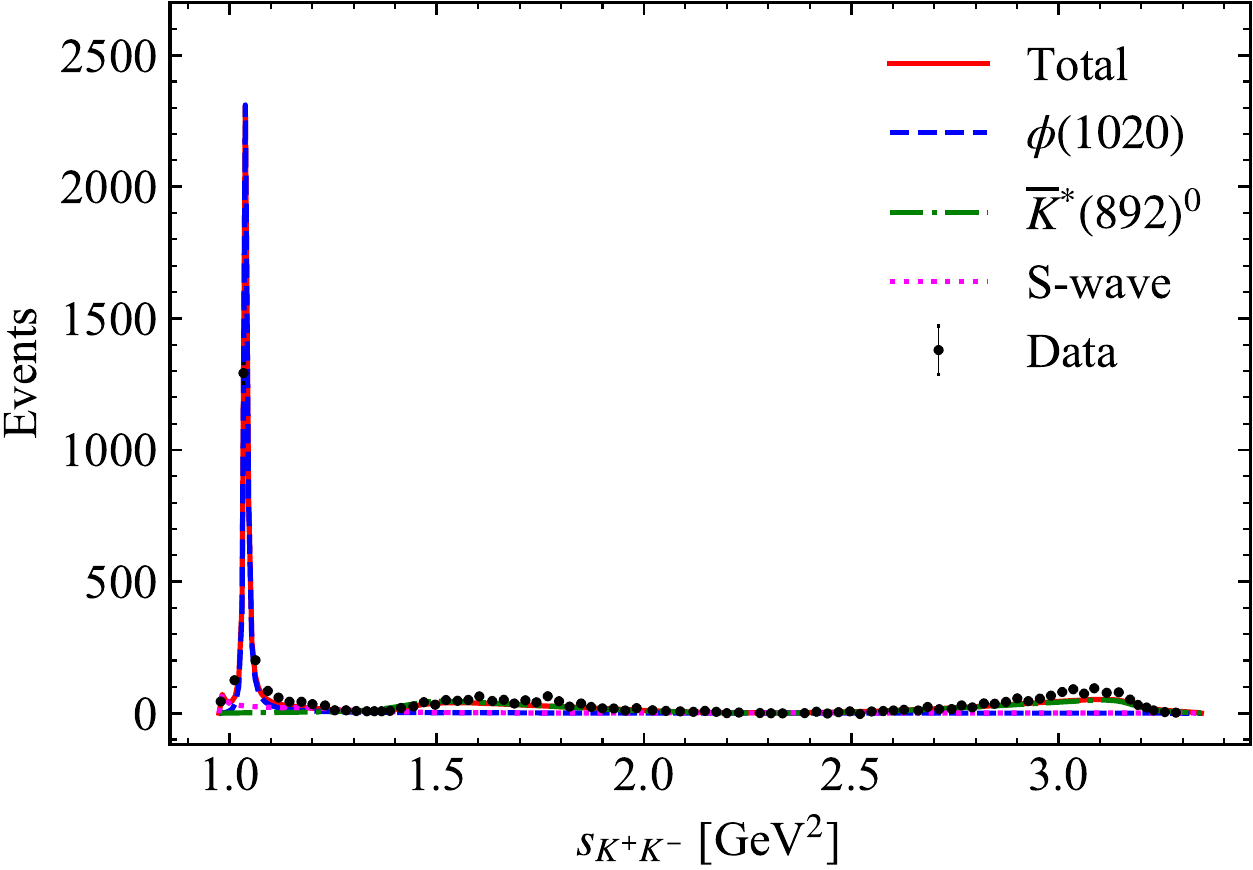} 
		\caption{\footnotesize $s_{K^{+}K^{-}}$}
		\label{fig:S12}
	\end{subfigure}
	\quad
	\begin{subfigure}{0.47\textwidth}  
		\centering 
		\includegraphics[width=1\linewidth]{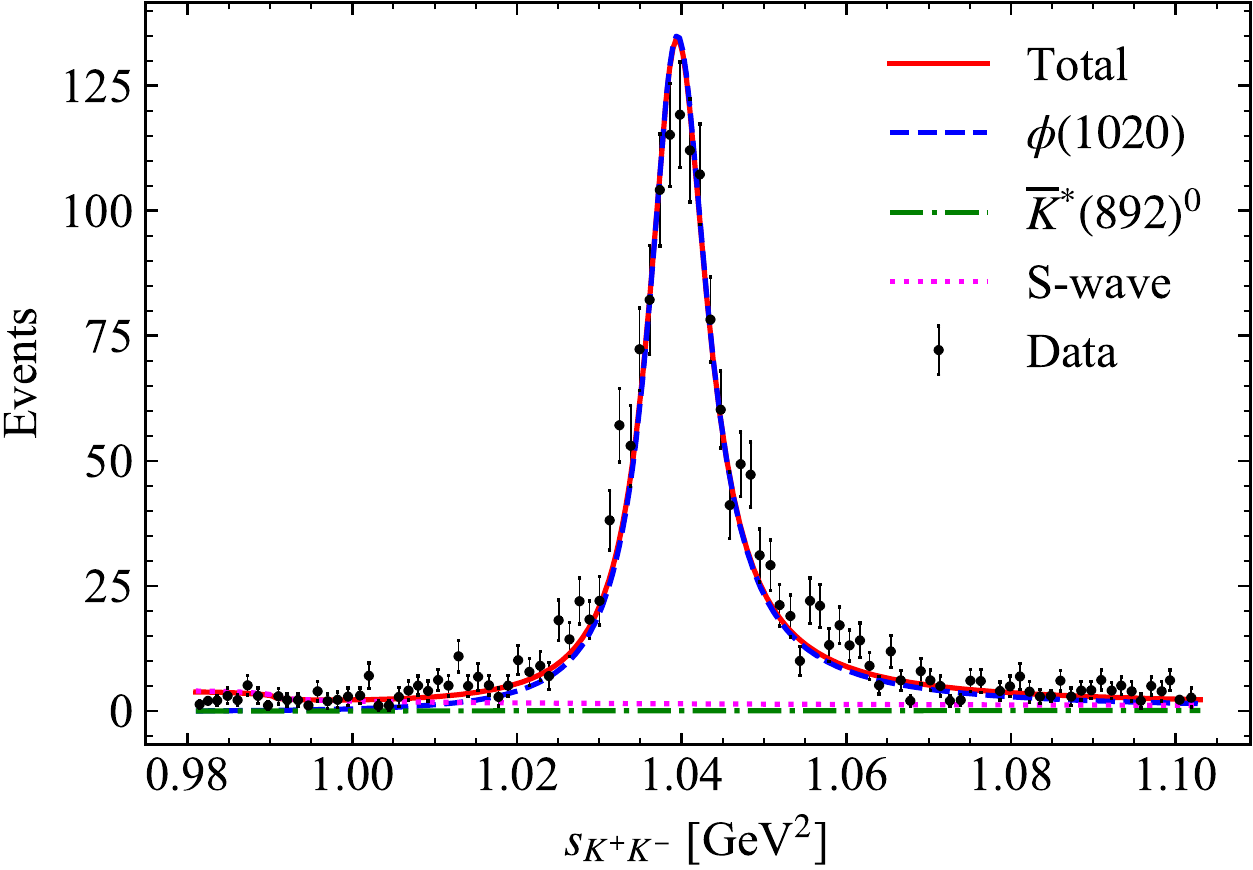} 
		\caption{\footnotesize $s_{K^{+}K^{-}}$ near the $\phi(1020)$ peak.}
		\label{fig:S122}  
	\end{subfigure}	
	\quad
	\begin{subfigure}{0.47\textwidth}  
		\centering
		\includegraphics[width=1\linewidth]{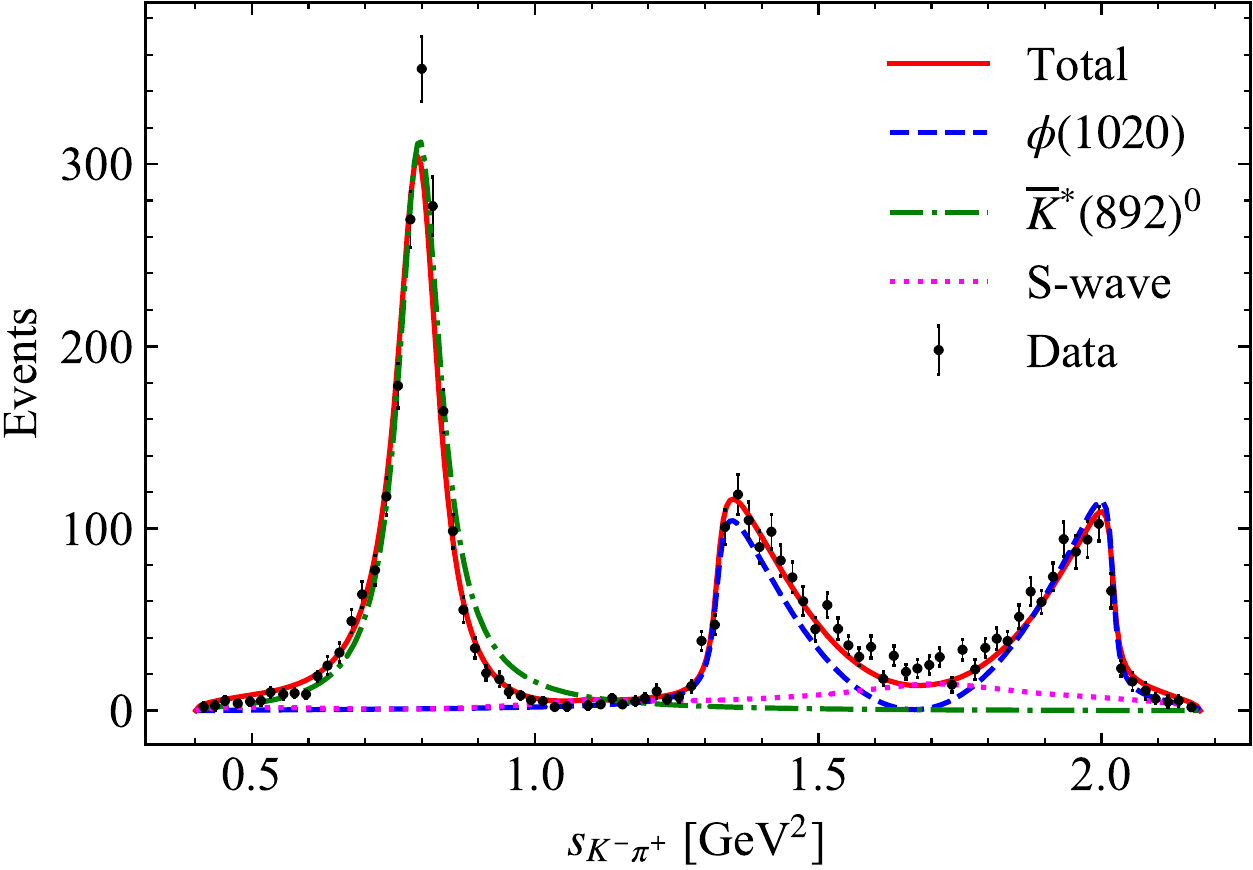} 
		\caption{\footnotesize $s_{K^{-}\pi^{+}}$ ($\chi^{2}/dof.=192.89/(87-7)=2.41$)}
		\label{fig:S23}
	\end{subfigure}
	\quad
	\begin{subfigure}{0.47\textwidth}  
		\centering 
		\includegraphics[width=1\linewidth]{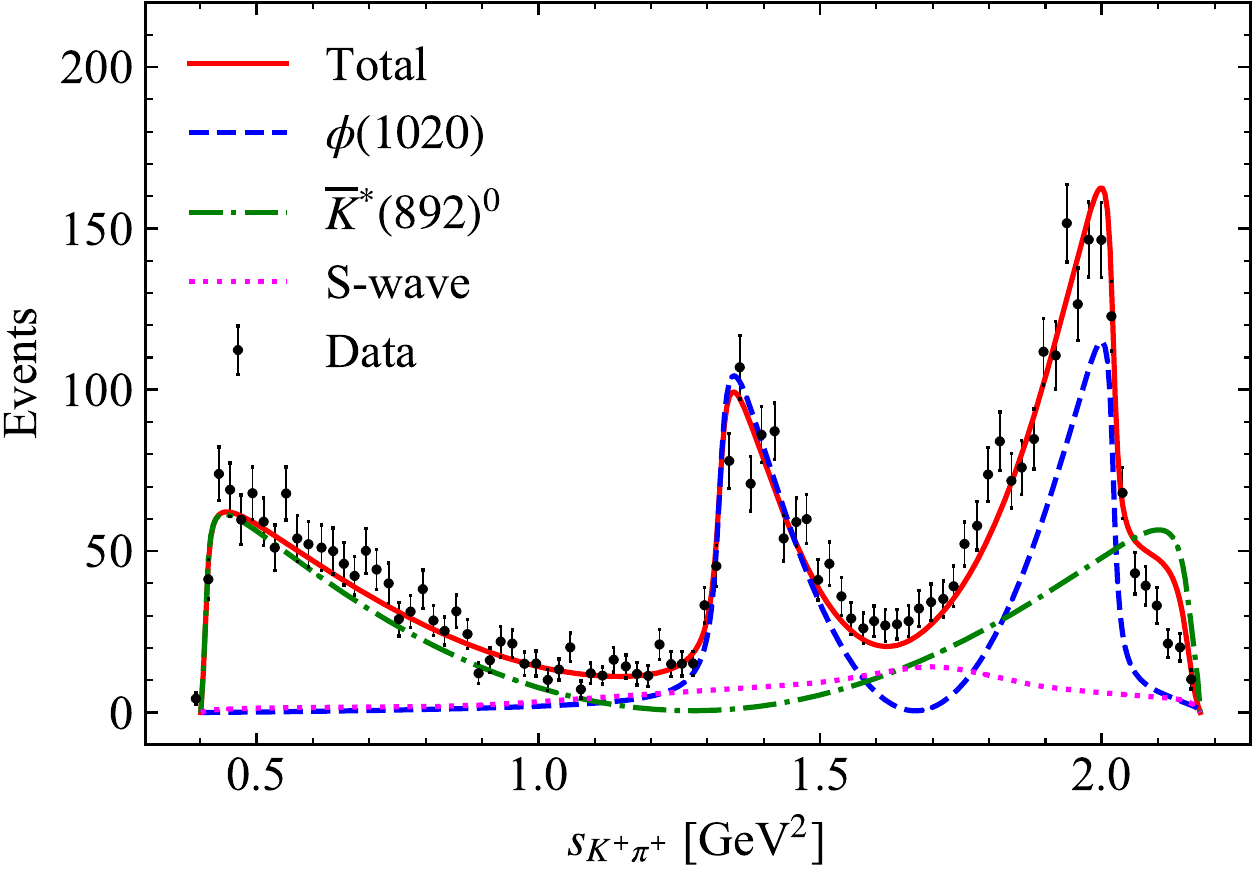} 
		\caption{\footnotesize $s_{K^{+}\pi^{+}}$}
		\label{fig:S13}  
	\end{subfigure}		 
	\caption{Fitting for different invariant mass distributions of Dalitz plot projections data~\cite{Ablikim:2020xlq}. The solid (red) line is the total contributions of $S$- and $P$-waves, the dash (blue) line the contribution of $\phi(1020)$, the dash-dot (green) line the one of $\bar{K}^*(892)^{0}$, the dot (magenta) line the one from $S$-wave (mainly $f_0(980)$), and dot (black) points are the experimental data taken from Ref.~\cite{Ablikim:2020xlq}.}
	\label{fig:S}
\end{figure} 

\begin{figure}[htbp]  
 	\centering 
 	\includegraphics[width=0.6\linewidth]{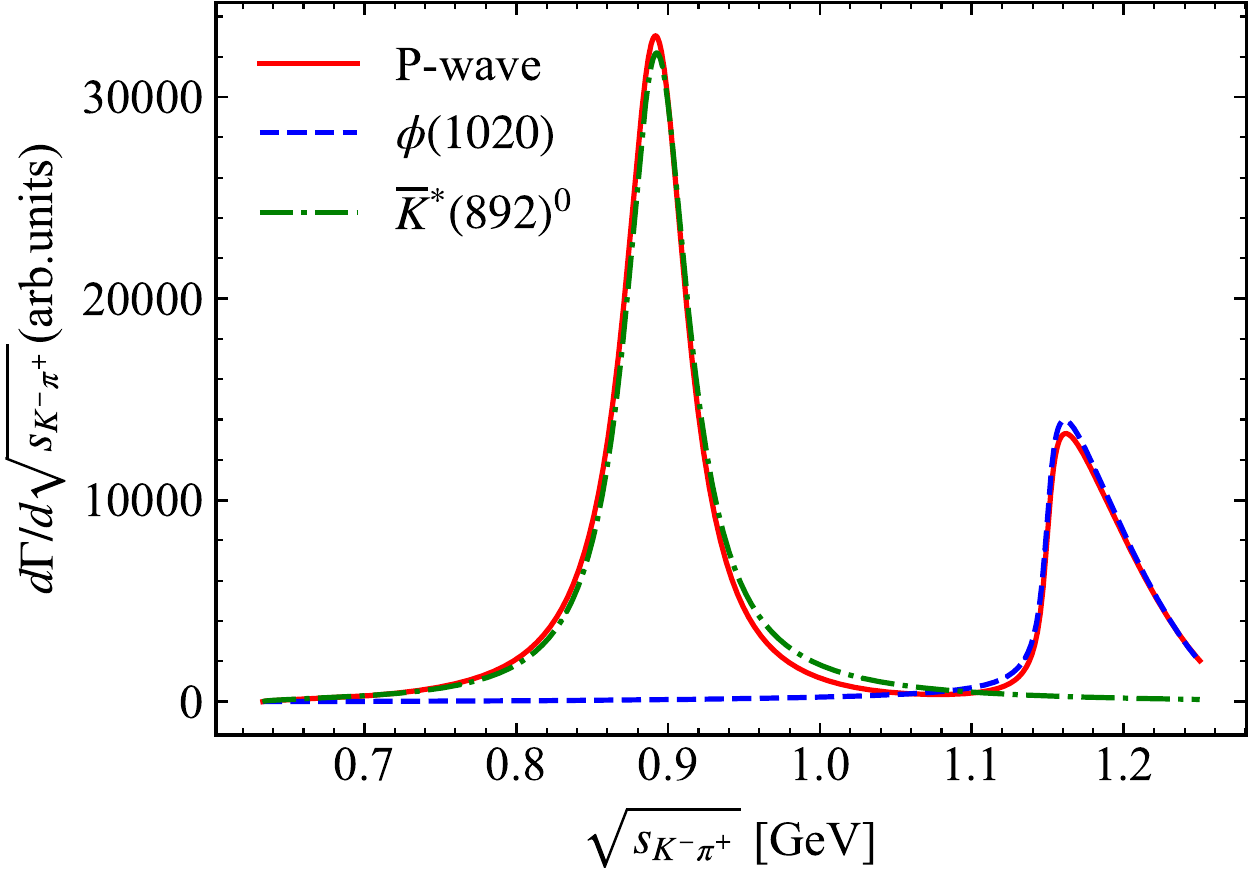} 
 	\caption{$K^{-} \pi^{+}$ invariant mass distribution in $P$-wave of the $D_{s}^{+} \rightarrow K^{+} K^{-} \pi^{+}$ decay.}
 	\label{fig:PMinv23}  
\end{figure}

Next, we perform a fit to the Dalitz plot projections data~\cite{Ablikim:2020xlq} for different invariant mass distributions, which is shown in Fig.~\ref{fig:S23}. The parameters from the fit are shown in Fit~\uppercase\expandafter{\romannumeral2} of Table~\ref{tab:Parameters}. Because there is only one experimental data for the peak near $1.02$ GeV in Fig.~\ref{fig:S12}, and the bins of the data in Fig. \ref{fig:S122} are different from those of the other three figures, we choose the data of $s_{K^{-}\pi^{+}}$ in Fig. \ref{fig:S23} to do the fit
\footnote{Note that we have done a fit with the combined data of $s_{K^{-}\pi^{+}}$ and $s_{K^{+}\pi^{+}}$ (in Figs.~\ref{fig:S23} and~\ref{fig:S13}), and found that there is no significant improvement to the fit only with the $s_{K^{-}\pi^{+}}$ data.}.
With the fitting parameters obtained, we can directly get the results of Figs.~\ref{fig:S12} and \ref{fig:S13}, except for Fig.~\ref{fig:S122}. Since the sampling interval and the number of events in Fig.~\ref{fig:S122} are different from the ones in Fig.~\ref{fig:S23}, apart from using the fitting parameters obtained, we add a global factor for the overall strength of the curve to match the different event numbers in Fig.~\ref{fig:S122}, of which the value is $0.054$. From Fig.~\ref{fig:S}, one can see that our fitting results are in good agreement with the experimental data. It is remarkable that we only fit the data of Fig.~\ref{fig:S23} and obtain good description of the other data in Figs.~\ref{fig:S12}, \ref{fig:S122} and \ref{fig:S13}, which is analogous to the case of $D^+\to K^-K^+K^+$ decay discussed in Ref.~\cite{Roca:2020lyi}. In Fig.~\ref{fig:S12}, the contribution of $\phi(1020)$ is obvious, and the two small bumps in the middle and high energy regions are contributed by $\bar{K}^{*}(892)^{0}$, see dash-dot (green) line. Fig.~\ref{fig:S122} shows the detailed structure of the $\phi(1020)$ state. For the $K^{-}\pi^{+}$ invariant mass distribution in Fig.~\ref{fig:S23}, besides the contribution of $\bar{K}^{*}(892)^{0}$, the two peaks in the middle and high energy regions are dominantly contributed by the $\phi(1020)$, see dash (blue) line. Similarly, for the $K^{+}\pi^{+}$ invariant mass distribution in Fig.~\ref{fig:S13}, the lower peak near $0.5$ GeV is contributed by $\bar{K}^{*}(892)^{0}$, and the other two are mainly contributed by the $\phi(1020)$ too. Note that we only put two resonances' contribution, the $\bar{K}^*(892)^{0}$ and $\phi(1020)$ states added by hand, and obtain the results of Figs.~\ref{fig:S23} and~\ref{fig:S13} in good agreement with experimental data rather than considering more higher resonances' contribution in the experimental analysis~\cite{Ablikim:2020xlq}. This is a curious result of our work, which is similar to the one obtained in Ref.~\cite{Toledo:2020zxj} for the investigation of $D^0 \to K^- \pi^+ \eta$ decay with only two resonances' contribution describing the data well. Furthermore, compared with the $P$-wave, the contribution of $S$-wave to Dalitz plot projection data is smaller, which is mainly concentrated in the $K^{-}\pi^{+}$ and $K^{+}\pi^{+}$ channels, see the dot (magenta) line in Fig.~\ref{fig:S}. Thus, there is no clear signal for the $K^*_0(700)$ resonance in $S$-wave, as found in the experiment~\cite{Ablikim:2020xlq}. Besides, it is obvious that the contribution of $\bar{K}^{*}(892)^{0}$ is very small compared with $\phi(1020)$ in the $K^{+} K^{-}$ invariant mass distribution, see Figs.~\ref{fig:PMinv12} and~\ref{fig:S122}. Thus, the fitting parameters $\mathcal D_{\bar{K}^{*}}$ and $\alpha_{\bar{K}^{*}}$ may have large uncertainties in Fit~\uppercase\expandafter{\romannumeral1} of Table~\ref{tab:Parameters}, which is analogous to the one of $\mathcal D$ in Fit~\uppercase\expandafter{\romannumeral2} of Table~\ref{tab:Parameters}. Using the $P$-wave fitting parameters of Fit \uppercase\expandafter{\romannumeral2}, and the ratio of strength parameter $\mathcal D_{\phi}$ of $\phi(1020)$ between Fit \uppercase\expandafter{\romannumeral1} and Fit \uppercase\expandafter{\romannumeral2}, $1201.86/158.32=7.59$, we can easily get the corresponding $K^{-}\pi^{+}$ invariant mass distribution, which is shown in Fig.~\ref{fig:PMinv23} and can be used to evaluate the branching ratio below.  

In addition, we make a calculation for the ratios of branching fractions in different decay channels. By integrating the resonance contributions in $D_{s}^{+} \rightarrow K^{+} K^{-} \pi^{+}$ decays (recall that there is no contribution of the $a_{0}(980)$ state), we find
\begin{equation} 
	\begin{aligned}
		\frac{\mathcal{B}[D_{s}^{+} \rightarrow f_{0}(980) \pi^{+}, f_{0}(980) \rightarrow K^{+} K^{-}]}{\mathcal{B}[D_{s}^{+} \rightarrow \phi(1020) \pi^{+}, \phi(1020) \rightarrow K^{+} K^{-}]} = 0.28 ^{+0.03}_{-0.08}\; ,
	\end{aligned}
	\label{fig:Ratios1}
\end{equation} 
where the integral limit is $[2m_{K},1.1]$ GeV for both $D_{s}^{+} \rightarrow f_{0}(980) \pi^{+}$ and $D_{s}^{+} \rightarrow \phi(1020) \pi^{+}$ decays, with the uncertainty from the upper limits $1.1 \pm 0.05 $ GeV. Analogously, we get
\begin{equation} 
	\begin{aligned}
		\frac{\mathcal{B}[D_{s}^{+} \rightarrow \bar{K}^{*}(892)^{0} K^{+}, \bar{K}^{*}(892)^{0}\rightarrow K^{-} \pi^{+}]}{\mathcal{B}[D_{s}^{+} \rightarrow \phi(1020) \pi^{+}, \phi(1020) \rightarrow K^{+} K^{-}]} = 1.18^{+0.02}_{-0.05}\; ,
	\end{aligned}
	\label{fig:Ratios2}
\end{equation} 
where the integral limit is $[(m_{K}+m_{\pi}),1.1]$ GeV for $D_{s}^{+} \rightarrow  \bar{K}^{*}(892)^{0} K^{+}$ decay, and for the contributions of $\phi(1020)$ and $\bar{K}^{*}(892)^{0}$, we integrate the $P$-wave contribution of dash (blue) line in Fig.~\ref{fig:PMinv12} and dash-dot (green) line in Fig.~\ref{fig:PMinv23}, respectively. Taking the experimental measurement of the branching fraction from BESIII Collaboration~\cite{Ablikim:2020xlq} $\mathcal{B} [D_{s}^{+} \rightarrow \phi(1020) \pi^{+}, \phi(1020) \rightarrow K^{+} K^{-}]=(2.21\pm0.05\pm0.07)\%$ as input, and combining Eqs.~\eqref{fig:Ratios1} and \eqref{fig:Ratios2}, one can easily obtain the branching fractions,
\begin{equation} 
	\begin{aligned}
		\mathcal{B}[D_{s}^{+} \rightarrow f_{0}(980) \pi^{+}, f_{0}(980) \rightarrow K^{+} K^{-}]= (0.61 \pm 0.02 ^{+0.06}_{-0.17})\,\%, \\
		\mathcal{B}[D_{s}^{+} \rightarrow \bar{K}^{*}(892)^{0} K^{+}, \bar{K}^{*}(892)^{0} \rightarrow K^{-} \pi^{+}]= (2.61 \pm 0.10 ^{+0.05}_{-0.12})\,\%,
	\end{aligned}
	\label{fig:Theory}
\end{equation}
where the first uncertainties are estimated from the experimental error of $\mathcal{B} [ D_{s}^{+} \rightarrow \phi(1020) \pi^{+},\ \\ \phi(1020) \rightarrow K^{+} K^{-} ]$, and the second ones come from the integration limits of Eqs. (\ref{fig:Ratios1}) and (\ref{fig:Ratios2}). Note that the corresponding branching ratios from BESIII Collaboration were reported as~\cite{Ablikim:2020xlq}
\begin{equation} 
	\begin{aligned}
		\mathcal{B}[D_{s}^{+} \rightarrow S(980) \pi^{+}, S(980) \rightarrow K^{+} K^{-}]= (1.05\pm0.04\pm0.06)\%, \\
		\mathcal{B}[D_{s}^{+} \rightarrow \bar{K}^{*}(892)^{0} K^{+}, \bar{K}^{*}(892)^{0}\rightarrow K^{-} \pi^{+}]= (2.64\pm0.06\pm0.07)\%,
	\end{aligned}
	\label{fig:BESIII}
\end{equation}
where $S(980)$ represents the states of $f_{0}(980)$ and $a_{0}(980)$. Moreover, the ones reported in PDG~\cite{Zyla:2020pdg} are given by,
\begin{equation} 
	\begin{aligned}
		\mathcal{B}[D_{s}^{+} \rightarrow f_{0}(980) \pi^{+}, f_{0}(980) \rightarrow K^{+} K^{-}]= (1.14\pm0.31)\%, \\
		\mathcal{B}[D_{s}^{+} \rightarrow \bar{K}^{*}(892)^{0} K^{+}, \bar{K}^{*}(892)^{0}\rightarrow K^{-} \pi^{+}]=(2.58\pm0.08)\%.
	\end{aligned}
	\label{fig:BABAR}
\end{equation}
One can see that our branching fraction of $\mathcal{B}[D_{s}^{+} \rightarrow \bar{K}^{*}(892)^{0} K^{+}, \bar{K}^{*}(892)^{0}\rightarrow K^{-} \pi^{+}]$ is consistent with the one obtained in the experimental results of Ref.~\cite{Ablikim:2020xlq} and PDG~\cite{Zyla:2020pdg} within the uncertainties. Whereas, the one of $\mathcal{B}[D_{s}^{+} \rightarrow f_{0}(980) \pi^{+}, f_{0}(980) \rightarrow K^{+} K^{-}]$ is about 40\% smaller than theirs.

\section{Conclusions}
\label{sec:Conclusions}

Based on the chiral unitary method for two-body final state interactions, we investigate the $D_{s}^{+} \rightarrow K^{+} K^{-} \pi^{+}$ decay with the final state interaction approach. In the decay process of hadronization, we have considered the mechanisms of external and internal $W$-emission. When we sum all the contributions from the final states in $S$-wave, we find only $f_0(980)$ with isospin $I=0$ contributed and without the one of $a_{0}(980)$ as indicated in both the experiment~\cite{delAmoSanchez:2010yp} and theories~\cite{Dias:2016gou,Wang:2021naf}. Note that the contribution of $f_0(980)$ is produced in the interactions of $K^+ K^-$ with its coupled channels, which means that the $f_0(980)$ is a bound state of $K\bar{K}$ and locates below the threshold of $K\bar{K}$. On the other hand, as found in our fitting results, the contribution from $S$-wave is small. This is why the contribution from $f_0(980)$ or/and $a_{0}(980)$ could not be distinguished in recent experimental analysis~\cite{Ablikim:2020xlq}. 

With only seven free parameters and three resonances' contribution in both $S$- and $P$-waves, we obtain the results of the invariant mass distribution in good agreement with the experimental data. It is remarkable that only taking into account three resonances' contribution, the $f_0(980)$, $\bar{K}^{*}(892)^{0}$ and $\phi(1020)$ states, the experimental invariant mass distributions of $K^+ K^-$, $K^+ \pi^+$ and $K^- \pi^+$ can be described well without any higher resonances' contribution as done in the experimental analysis~\cite{Ablikim:2020xlq}. The other feature of our results is that only fitting with the $K^- \pi^+$ invariant mass distribution, one can get good description of other invariant mass distributions of the Dalitz plot projection data. As shown in these fitting results, for the $K^{+} K^{-}$ invariant mass distribution, except of the clear $\phi(1020)$ peak, the two small bumps in the middle and high energy regions are caused by the $\bar{K}^{*}(892)^{0}$. For the $K^+ \pi^+$ and $K^- \pi^+$ invariant mass distributions, the lower peak is mainly contributed by the $\bar{K}^{*}(892)^{0}$, while the other two peaks in the middle and high energy regions are dominated by the $\phi(1020)$. Furthermore, we also calculate the branching fractions of the dominant decay channels with the scalar and vector resonances produced in the final states. We obtain the result of the branching ratio of $\mathcal{B}[D_{s}^{+} \rightarrow \bar{K}^{*}(892)^{0} K^{+}, \bar{K}^{*}(892)^{0}\rightarrow K^{-} \pi^{+}] = (2.61 \pm 0.10 ^{+0.05}_{-0.12})\,\%$, which is consistent with the experimental measurement from BESIII Collaboration and Particle Data Group within the uncertainties, and the one of $\mathcal{B}[D_{s}^{+} \rightarrow f_{0}(980) \pi^{+}, f_{0}(980) \rightarrow K^{+} K^{-}] = (0.61 \pm 0.02 ^{+0.06}_{-0.17})\,\%$ a bit smaller.

\section*{Acknowledgments}

We thank Prof. Eulogio Oset for careful reading the manuscript and valuable comments, and acknowledge Profs. Bastian Kubis, Yu-Kuo Hsiao for helpful comments. ZFS is partly supported by the National Natural Science Founadtion of China (NSFC) under Grants No. 11965016, and the Fundamental Research Funds for the Central Universities under Grants No. Lzujbky-2021-sp24.

 \addcontentsline{toc}{section}{References}
\end{document}